\documentclass[12pt]{article}

\usepackage{graphics}
\usepackage{subfigure}
\usepackage{booktabs}
\usepackage{amsfonts,amssymb}
\usepackage[figuresright]{rotating}
\usepackage{bm}
\usepackage{epsf}
\usepackage{epsfig}

\newcommand{\be}{\begin{equation}}
\newcommand{\ee}{\end{equation}}
\newcommand{\bc}{\begin{center}}
\newcommand{\ec}{\end{center}}

\begin{document}

{\normalsize \hfill ITP-UU-08/35}\\
\vspace{-1.5cm}
{\normalsize \hfill SPIN-08/26}\\
${}$\\

\bc
\vspace{48pt}
{ \Large \bf Shaken, but not stirred -- \\
{}${}$\\
Potts model coupled to quantum gravity}

\vspace{50pt}

{\sl J. Ambj\o rn}$\,^{a,c}$,
{\sl K.N. Anagnostopoulos}$\,^{b}$,
{\sl R. Loll}$\,^{c}$,
{\sl I. Pushkina}$\,^{c}$

\vspace{24pt}
{\footnotesize

$^a$~The Niels Bohr Institute, Copenhagen University \\
Blegdamsvej 17, DK-2100 Copenhagen \O, Denmark \\
{email: ambjorn@nbi.dk}\\

\vspace{10pt}

$^b$~Physics Department, National Technical University of Athens \\
Zografou Campus, GR-15780 Athens, Greece \\
{email: konstant@mail.ntua.gr}

\vspace{10pt}

$^c$~Institute for Theoretical Physics, Utrecht University \\
Leuvenlaan 4, NL-3584 CE Utrecht, The Netherlands \\
{email: r.loll@phys.uu.nl, i.pushkina@phys.uu.nl} \\

\vspace{10pt}
}
\vspace{20pt}

{\sl 19 Jun 2008}

\vspace{48pt}

\ec

\bc
{\bf Abstract}
\ec
We investigate the critical behaviour of both matter and geometry 
of the three-state Potts model coupled to
two-dimensional Lorentzian quantum gravity in the framework of causal dynamical triangulations. 
Contrary to what general arguments on the effects of disorder suggest, we find strong numerical evidence that 
the critical exponents of the matter are not changed under the influence of quantum fluctuations in the geometry, compared to their values on fixed, regular lattices. 
This lends further support to previous findings that quantum gravity models based on causal dynamical triangulations are in many ways better behaved than their Euclidean counterparts.

\thispagestyle{empty}

\newpage
\setcounter{page}{1}

\section{Matter and geometry in two dimensions}\label{intro}

A common difficulty for models of nonperturbative quantum gravity, which attempt to describe a Planckian regime of quantum-fluctuating and strongly coupled degrees of freedom, is to reproduce aspects of the classical theory of general relativity in a suitable large-scale limit. Assuming that one has a quantum model which is sufficiently complete to be considered a candidate for a quantum gravity theory, it will by construction not be given in terms of small metric fluctuations around some classical spacetime geometry. It is then a nontrivial step to show that classical gravity does indeed emerge on larger scales, and to elucidate the mechanism by which this happens. 

One way to probe the properties of (quantum) geometry is by coupling matter to the gravitational system and observing its behaviour as a function of scale.
A necessary condition for the existence of a good classical limit is that on sufficiently large scales and for sufficiently weak gravity the matter should behave like on a fixed, classical background geometry. That this should happen is by no means obvious, if one starts from a Hilbert space or a path integral encoding nonperturbative Planck-scale excitations. The latter may be too numerous or violent to coalesce into a well-behaved macroscopic, four-dimensional spacetime. This kind of pathological behaviour is not just an abstract possibility, but has been found in Euclidean models of quantum gravity \cite{4deu}, and exhibits a certain genericity.    

The set of spacetime geometries\footnote{that is, the ``histories" contributing to the ``sum over histories", a.k.a. the
gravitational path integral} underlying the approach of Causal Dynamical Triangulations (CDT) to quantum gravity seems to strike a balance between generating large quantum excitations on small scales \cite{ajl-rec}, and managing to reproduce features of classical geometry on large scales \cite{ajl-prl,agjl}. Something similar is apparently true for the {\it two}-dimensional quantum gravity theory derived from a CDT formulation \cite{al,2d,ajwz,alwz}. Although there is no classical limit per se for the purely gravitational degrees of freedom in this case -- the Einstein-Hilbert action is trivial -- one can still ask whether the dynamics of any additional matter fields is changed on a quantum-fluctuating ``background geometry", compared with the same matter on a fixed background. For the much-studied case of two-dimensional Euclidean (``Liouville") quantum gravity, obtained from Euclidean dynamical triangulations (EDT) or their equivalent matrix models  \cite{liouville}, one indeed finds that gravity alters the matter behaviour nontrivially. Namely, for matter with a central charge $c$, $0<c\leq 1$, its critical exponents appear   
as ``dressed" versions of their fixed-background counterparts. While these models present interesting examples of strongly interacting ``gravity"-matter systems, their existence is restricted by the so-called $c=1$-barrier for the central charge, beyond which no consistent uni\-tary matter models have been found. A closer inspection of the geometric properties of these models suggests that beyond the barrier, the backreaction of matter on geometry is too strong to be compatible with an extended, connected carrier space; space simply shrivels to a branched polymer \cite{ad,david}.    

\begin{figure}[hbt]
 \centering
 \includegraphics[width=1.\linewidth]{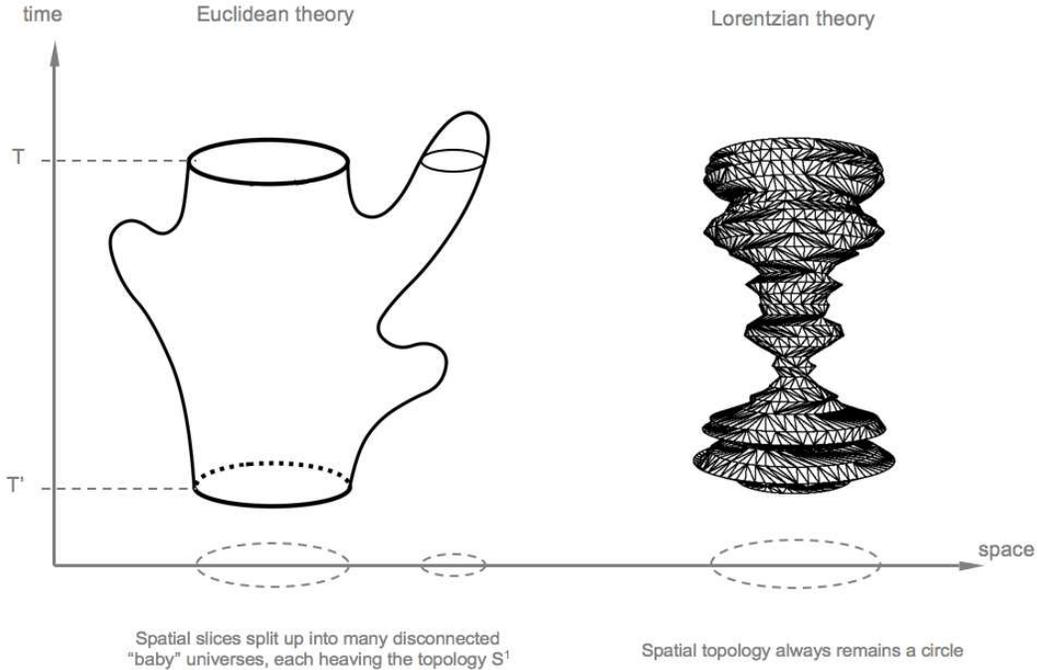}
 \caption{\small Illustrating the difference between two-dimensional path integral histories of Euclidean and Lorentzian signature, both of cylinder topology. Superimposing a proper-time slicing onto a two-dimensional Euclidean space (left), starting from an arbitrary initial circular slice will generically lead to disconnected slices of topology $S^1\times S^1\times ...$ at later times. By definition, such ``baby universes" are absent from the Lorentzian spacetimes (right), where the spatial topology always remains a single $S^1$.}
 \label{fig1}
\end{figure}

By contrast, the indications so far from analogous two-dimensional models of {\it causal, Lorentzian} gravity coupled to Ising spins are that (i) the universal properties of the matter are not altered by the quantum fluctuations of the ensemble of CDT geometries \cite{1q2LQG}, and (ii) the models remain consistent beyond the $c=1$-barrier, where the latter comes from a numerical study of coupling to multiple (in this case, eight) copies of Ising spins \cite{8q2LQG}. In this sense, Lorentzian quantum gravity, based on a path integral over causal spacetime geometries, also here turns out to be better behaved than its Euclidean counterpart, based on a path integral over isotropic ``spacetimes", which make no distinction between time-like and spatial directions, see Fig.\ \ref{fig1}.

While coupled quantum systems of matter and geometry are clearly of central interest to practitioners of nonperturbative, four-dimensional quantum gravity, for which lower-dimensional systems may serve as useful toy models, they can also be considered from a quite different angle, that of statistical mechanical systems with {\it disorder}, and their associated critical behaviour as a function of the strength of the disorder. In the context of spin models and, more specifically, the $q$-state ferromagnetic Potts models, which we shall concentrate on in what follows, one often looks at {\it bond} disorder. The role of random variables in this case is played by the local spin couplings $J_{ij}\geq 0$ in the energy $-\sum_{\langle ij\rangle} J_{ij} \delta_{\sigma_i \sigma_j}$, where the sum is taken over all nearest neighbours $i$ and $j$ on a fixed, regular lattice, and the spins take values $\sigma_i=1, ..., q$. For a discussion of the (somewhat controversial) state of the art with regard to the role of quenched bond disorder, see 
\cite{jajo} and references therein.

Instead, what we will focus on in the present work is a geometric type of disorder, also termed {\it connectivity disorder}, in reference to the irregular nature of the lattice geometry. The situation we have in mind is that of two-dimensional curved spaces (of fixed topology) obtained from gluing together flat, equilateral triangles, precisely as in the models of dynamical triangulations.\footnote{Whenever the number of triangles meeting at a vertex -- its ``connectivity" or ``coordination number" -- is not equal to six, the vertex carries a nontrivial intrinsic curvature.} The quantum gravity context implies that we are primarily interested in so-called {\it annealed} disorder, where an ensemble average is taken over the disorder parameter, in our case, in the form of a sum over triangulated spacetime histories of a given discrete volume $N$. Unlike in models with {\it quenched} disorder, where one studies the spin system on a fixed, disordered lattice, the disorder in an annealed setting is itself part of the dynamics, which allows in particular for a backreaction of matter on geometry. Depending on the details of the matter-gravity interactions, this tends to increase the strength of the disorder, when compared to the quenched setting.

The evidence from models with annealed and quenched connectivity 
disorder induced by {\it Euclidean} quantum gravity is that their universal properties are altered.
The annealed case, obtained by coupling 
Potts models with $q=2, 3, 4$ (with corresponding central charges $c=1/2, 4/5, 1$)
to an ensemble of EDT, is rather clear-cut, with predictions from exact matrix-model solutions
well confirmed by numerical simulations (see, for example, \cite{baillie}). As can be seen by comparing the
critical exponents with those of the same Potts models on regular, flat lattices (see Table \ref{table1}), 
the transitions
are ``softened", that is, the index $\alpha$ characterizing the behaviour of the heat capacity at
criticality is lowered. Theoretical arguments \cite{johnston} predict a similar -- though less drastic -- effect for the corresponding quenched systems, with a set of non-rational exponents, but these have not so far been corroborated convincingly by
simulations \cite{jajo}, and their status remains unclear. Additional arguments for the presence of new, quenched exponents (without predicting their actual values) come from adapting the Harris-Luck criterion \cite{harris,luck}, predicting the effect of disorder on the nature of phase transitions, to correlated geometric disorder \cite{jawei}.
\begin{table}[htb]
 \hspace{0.01\textwidth}
\begin{minipage}{0.4\textwidth}
 \centering
 \begin{tabular}{|c|c|c|c|}\toprule
 \multicolumn{4}{|c|}{\bf Potts models: fixed background} \\ \cmidrule(r){2-4}
  {\bf Indices} & {\bf 1q2}  & {\bf 1q3}  & {\bf 1q4} \\ \midrule
   $\alpha$     &     0      &   0.333    &   0.666   \\ 
   $\beta$      &   0.125    &   0.111    &   0.083   \\ 
   $\gamma$     &   1.75     &   1.444    &   1.166   \\ 
    $\nu d_H$       &     2      &   1.667    &   1.333   \\ \bottomrule
 \end{tabular}
\end{minipage}
 \hspace{0.1\textwidth}
\begin{minipage}{0.4\textwidth}
 \centering
 \begin{tabular}{|c|c|c|c|c|}\toprule
 \multicolumn{4}{|c|}{\bf Potts models coupled to EQG}     \\ \cmidrule(r){2-4}
  {\bf Indices} &  {\bf 1q2}  & {\bf 1q3}  & {\bf 1q4} \\ \midrule
    $\alpha$    &               -1      &    -0.5    & 0   \\ 
    $\beta$     &               0.5     &     0.5    & 0.5 \\ 
    $\gamma$    &              2      &     1.5    & 1   \\ 
     $\nu d_H$      &           3     &     2.5   & 2   \\
      \bottomrule
 \end{tabular}
\end{minipage}
\label{table1}
\caption{\small Comparing the critical exponents $\alpha$, $\beta$, $\gamma$ and
$\nu d_H$ of the $q$-state Potts models, $q=2,3,4$, on fixed, regular lattices (left) and 
coupled to Euclidean quantum gravity in the form of Euclidean Dynamical Triangulations (right). 
The notation $xqy$ refers to a model with $x$ copies of the $y$-state Potts model.}
\end{table}

Given the rich and only partially understood structure of the Euclidean case, one is clearly interested in getting a more complete picture of coupled matter-gravity models for {\it Lorentzian} gravity than is currently available, and what these models might teach us about the interplay of matter and geometry. 
The studies up to now have been limited to the Ising model, and were either numerical \cite{1q2LQG,8q2LQG} or involved a high-temperature expansion \cite{ising}. As already mentioned, both for a single copy of the Ising model (``1q2") and for eight copies thereof (``8q2"), strong evidence for the Onsager values for the matter exponents was found. This indicates a remarkable robustness of the ``flat-space" Onsager exponents, since the geometry is anything but flat -- it quantum-fluctuates and, in the 8q2 case, even changes its 
Hausdorff dimension $d_H$ from two to three \cite{8q2LQG}, signalling a phase transition of the geometric sector of the model.
 
In the present work, we will undertake a numerical analysis of the three-state Potts model, with $q=3$ and a central charge of 4/5, coupled to an ensemble of fluctuating geometries represented by CDTs. 
We will determine a number of critical exponents which characterize its universal behaviour, for both matter and geometry. 
This case is interesting for a number of reasons. With regard to disorder, it (like the CDT-Ising model) presumably lies in between the fixed, flat lattices and the annealed (as well as the quenched) Euclidean matter-coupled models, to which it can readily be compared. We are primarily interested in gathering further evidence or otherwise for the conjecture that matter on two-dimensional causal dynamical triangulations {\it always} behaves like on a fixed, flat lattice. If this was true in general and if one was just interested in extracting the universal properties of the matter model, it would open other interesting possibilities. First, from a numerical point of view, putting matter on ``flexible", fluctuating lattices may speed up the approach to the continuum limit, in the spirit of the old ``random lattice" program for field theory \cite{christ}. Some evidence for this for the case of the CDT-Ising model was found in \cite{ising}. Second -- and maybe surprisingly -- from the point of view of finding exact solutions, including a full sum over geometries {\it can} simplify this task, 
as demonstrated by the example of the EDT-Potts models, which have been solved by matrix-model techniques. Alas, since the order of the phase transitions in these instances is altered from the flat case, the solutions do not give us any new insights into solving the standard Potts model for $q>2$, which would be of great theoretical interest. If it is correct that these models on dynamical CDT lattices lie in the same universality class as those on fixed, flat lattices, this may now be coming within reach, in view of the fact that a matrix-model formulation for the pure CDT model in two dimensions has just been found \cite{matrix}.

Do we have any theoretical predictions for the behaviour of the 1q3 system? There is a relevance criterion for systems with quenched geometric disorder due to Janke and Weigel, who adapted and generalized the original Harris criterion to account for possible effects of long-range correlations among the disorder degrees of freedom \cite{jawei}, similar to what was done in \cite{luck} for quasi-crystals and other aperiodic lattices. The Harris criterion gives a threshold value $\alpha =0$ (for the system {\it without} disorder) for the specific-heat exponent, above which the disorder will be relevant \cite{harris}. 
According to \cite{jawei}, this threshold is lowered for sufficiently long-range correlations between the coordination numbers of the random lattice. In other words, in the presence of correlations, even more systems (namely, those with $\alpha >\alpha_c$,
$\alpha_c<0$) will change their universal behaviour as a result of the disorder. 
Assuming that the disorder is at most getting stronger in the annealed case, and without any
further information about the range of the correlations on the pure CDT lattices (which we have
not attempted to quantify), one would predict on the basis of these criteria that the critical behaviour of the three-state Potts model with $\alpha =1/3$ will be altered compared to the flat-lattice case. 
However, as we will describe in what follows, this is not what we find.

\section{Monte Carlo simulations}

We start by recalling that in the case of two-dimensional quantum gravity formulated in 
the framework of causal
dynamical triangulations, any curved spacetime is represented by gluing together identical, flat 
triangles\footnote{Since these simplices are by definition
flat on the inside, intrinsic curvature can only be located in places where more than two triangles meet, 
i.e. at the vertices.} with one space-like and two time-like edges. All basic triangular building blocks are identical
and their edge lengths fixed. The only degrees of freedom are therefore contained in the random way they are glued together pairwise along their edges. 
In contrast with Euclidean triangulations, where any gluing is allowed 
which gives rise to a piecewise flat manifold of a given topology, the gluing rules of the Lorentzian 
model are more restrictive and lead to two-dimensional spacetimes which are causal and possess
a global time arrow \cite{al}. This anisotropy with regard to space and time directions persists in the continuum limit, which can be computed analytically and gives rise to a new and inequivalent two-dimensional quantum gravity theory (see \cite{2d} for more details on the construction and solution of
the CDT pure-gravity model). 

After Wick rotation, the partition function of the three-state Potts model coupled to 
2d Lorentzian quantum gravity can be written as
\be
G(\lambda,t,\beta) = \sum_{T\in{\cal T}_t} e^{-\lambda N(T)} Z_T(\beta)
\ee
where the summation is over all triangulations $T$ of torus topology\footnote{that is, spacetimes 
with compact spatial $S^1$-slices, where (for reasons of simplicity and to minimize finite-size effects) time has also been cyclically identified} with $t$ time-slices, 
$N(T)$ counts the number 
of triangles in the triangulation $T$, and $\lambda$ is the bare cosmological constant. 
The matter partition function is given by
\be
Z_{T}(\beta) = \sum_{\{\sigma_i (T) \}} \exp\left( \beta\sum_{\langle ij\rangle \in T} \delta_{\sigma_i \sigma_j} \right),
\ee
where $\beta$ is proportional to the inverse temperature, $\beta=1/k\rm{T}$ (we have set $J_{ij}\equiv J=1$),
the spins take values $\sigma_i=0, 1, 2$ and $\langle ij\rangle$ denotes adjacent vertices in the appropriate class of causal triangulations $T$. We are
putting the matter spins at the vertices of the triangulation, but
could have also placed them at the triangle centres, 
since both choices are expected to lead to the same results in the scaling limit.
Recall that a geometry characterized by a toroidal triangulation $T$ of volume $N$ contains 
$N_v=N/2$ vertices, $N$ time-like links, $N/2$ space-like links, and thus $N_l=3N/2$ nearest-neighbour pairs in total.

We have simulated the Potts model for several lattices of sizes $N$ of between 200 and 125.000 triangles\footnote{All measurements presented were taken over the entire range of volumes, 
but are not presented in all the figures in order to avoid clutter (Figs.\ \ref{peaks}, \ref{fig:BC}).}, 
which on average have equal extension in time and spatial directions, that is, $N\simeq t^2$.
The Swendsen-Wang cluster algorithm was used to update the spin configuration (each sweep corresponding to approximately $N_v$
accepted moves), and typically about 500.000 measurements were taken at each $\beta$-value. 

\begin{figure}[hbt]
 \begin{center}
 \subfigure{\includegraphics[scale=0.40]{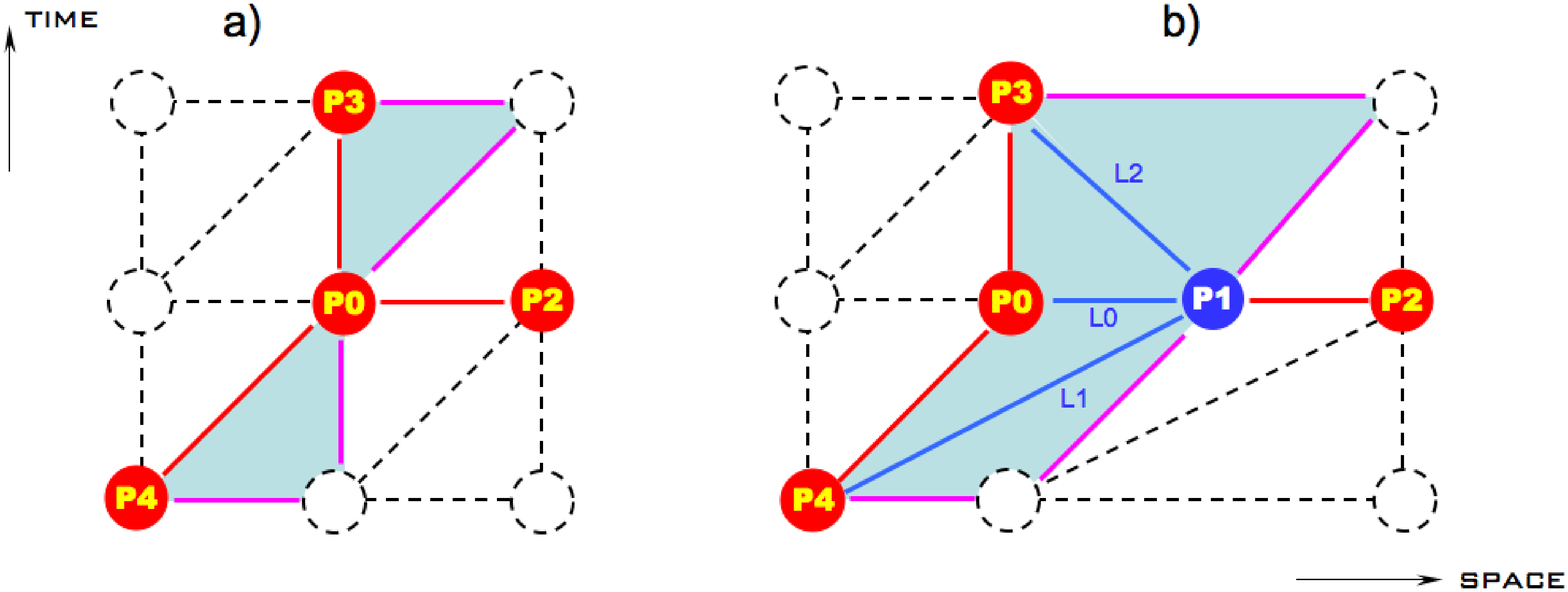}} \\
 \subfigure{\includegraphics[scale=0.40]{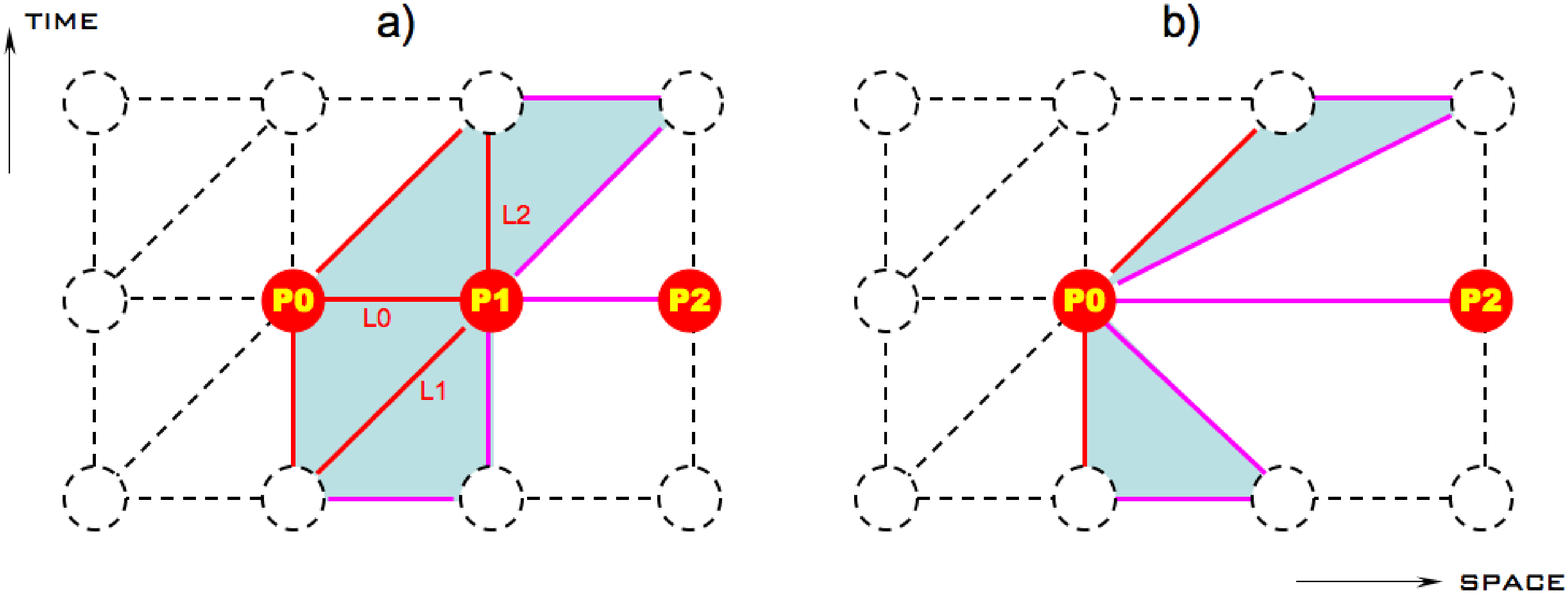}}
 \end{center}
 \caption{\small Local move changing the lattice geometry by insertion of a vertex (top), together with 
its inverse (bottom). Top: the move consists in adding a new vertex $P1$ to the right of a randomly chosen vertex $P0$ on the same spatial slice, together with three more links ($L0$, $L1$ and $L2$.) Bottom: the inverse move removes $P1$. 
The free edges are reconnected as if $P1$ had been slid on top of $P0$. 
}
\label{fig:moveAandB}
\end{figure}

\begin{figure}[tp]
 \subfigure{\includegraphics[scale=0.42]{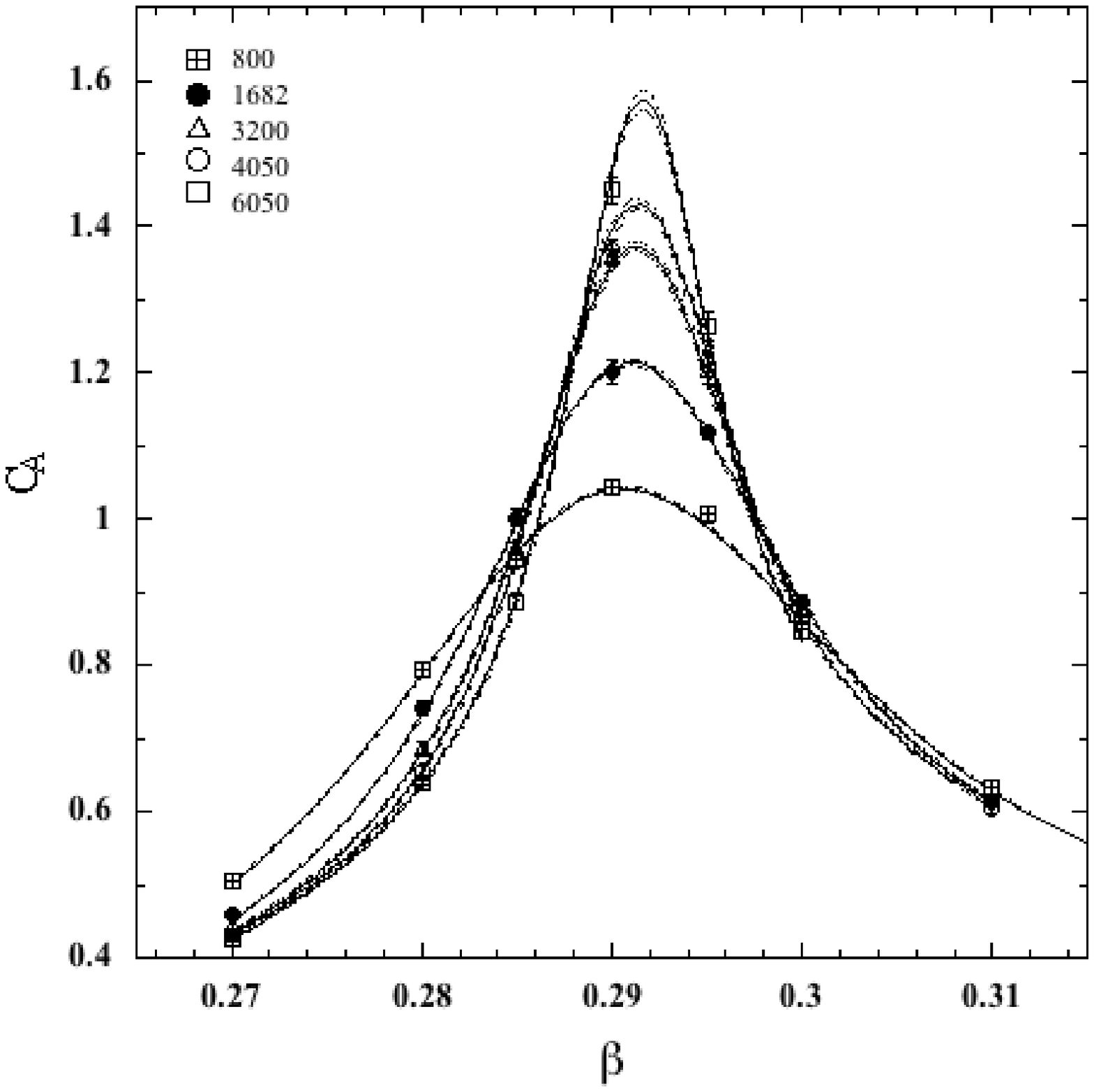}}
      \subfigure{\includegraphics[scale=0.42]{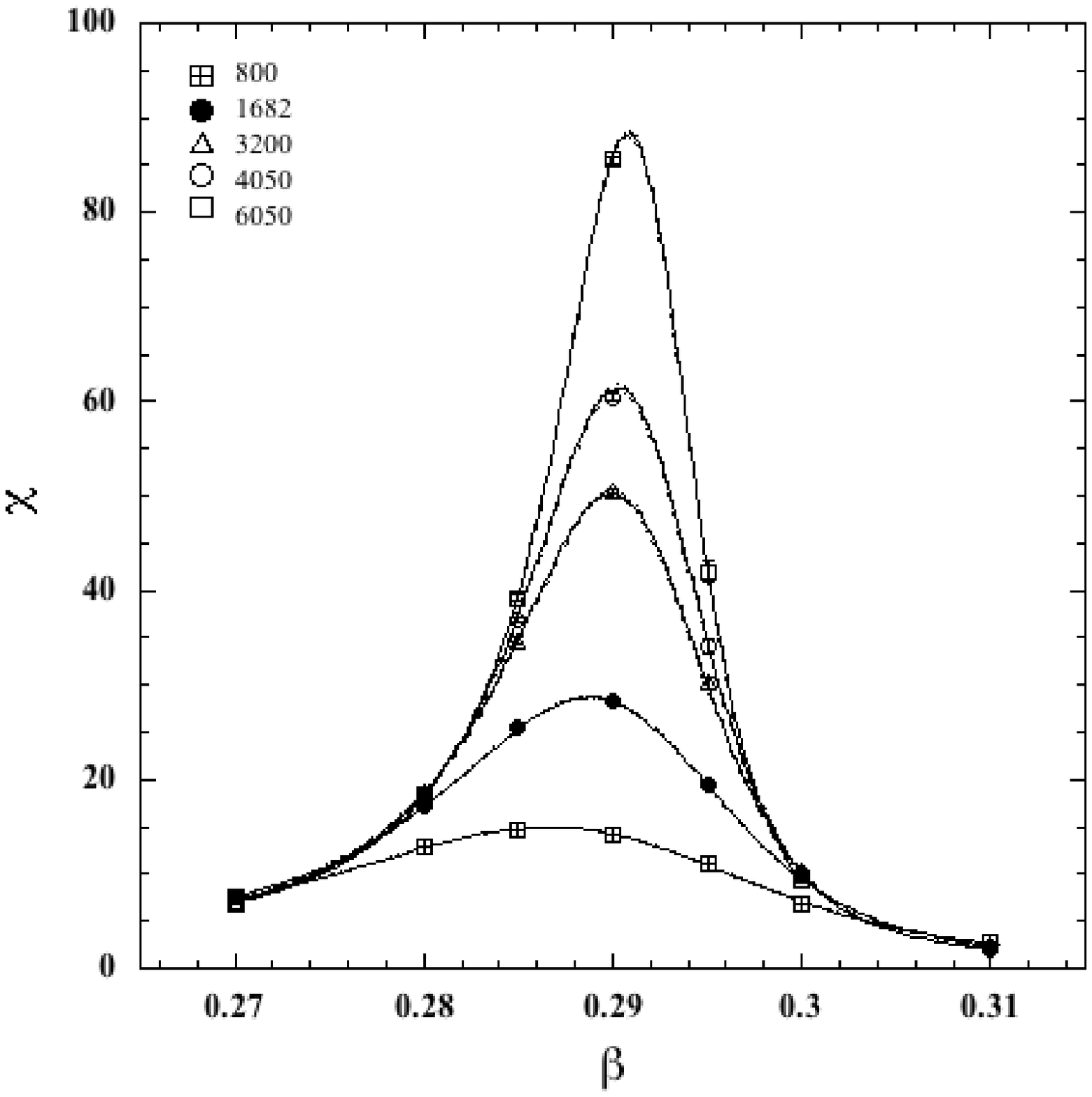}}
 \caption{\small The specific heat $C_A$ (left) and the magnetic susceptibility $\chi$ (right) of the 3-state Potts model coupled to Lorentzian quantum gravity, as function of the inverse temperature $\beta$, and for various volumes $N$.}
 \label{peaks}
\end{figure}

A local update of geometry involves a single {\it move} $A$ or its inverse $B$ 
(Fig.\ref{fig:moveAandB}), which together are ergodic in
the set of triangulations of a fixed number $t$ of time slices \cite{3d4d}. It consists in the insertion of a pair of triangles which share a spacelike link (one that lies entirely in a slice of constant time).
The geometry update is accompanied by an update of the spin configuration, and transition 
probabilities are assigned so as to satisfy detailed balance (see \cite{1q2LQG,DT-book}).

For each measurement (which is performed at fixed volume\footnote{Since the geometrical moves are not volume-preserving, in practice this is achieved by letting the volume fluctuate in a narrow band around its target value $N$, but only recording measurements collected at volume $N$ precisely.} $N$)
we stored the average energy density of the system per
link, $e=-\sum_{\langle ij\rangle }\delta_{\sigma_i \sigma_j}/N_l$ and the density per vertex $m$ of the (absolute value of the) magnetization. All observables we have studied in our analysis can be constructed from those two. 
Their scaling behaviour as a function of volume can be derived from standard finite-size scaling. 
They are
\begin{eqnarray}
&&\hspace{-1.2cm}\mbox{specific heat: } \hspace{1.8cm}
C_A = \beta^2 N\left(\langle e^2\rangle  - \langle e\rangle ^2\right) \sim  N^{\alpha/\nu d_H},
\label{alpha}\\
\vspace{.4cm}\nonumber\\
&&\hspace{-1.2cm}\mbox{magnetization: } \hspace{1.5cm}
m \sim N^{-\beta/\nu d_H}, \label{beta}\\
\vspace{.4cm}\nonumber\\
&&\hspace{-1.2cm}\mbox{magnetic susceptibility: }
\chi = N\left(\langle m^2\rangle  - \langle m\rangle ^2\right) \sim N^{\gamma/\nu d_H}.\label{gamma}
\end{eqnarray}
The critical exponents\footnote{We stick to the standard notation $\beta$ for both the inverse temperature and the critical exponent of the magnetization (the latter appearing in the exponent of
(\ref{beta}) and in relations (\ref{hyper})) -- this should not give rise to confusion.} 
 $\alpha$, $\beta$, $\gamma$ and $\nu$ characterize the behaviour of the 
system in the vicinity of the critical matter coupling $\beta_c$ (of the order-disorder transition), and are expected to 
satisfy the relations
\be
\alpha + 2\beta + \gamma = 2, \qquad 2\beta + \gamma = \nu d_H.
\label{hyper}
\ee
The parameter $\nu$ is the critical exponent of the divergent spin-spin correlation length and always appears in the combination $\nu d_H$ in the matter sector, where $d_H$ is the Hausdorff dimension. The latter characterizes the relation between linear distance $\ell$ (for example, the distance appearing in the correlator) and volume $V$ in the continuum limit according to $\ell \propto V^{1/d_H}$, for both $\ell$ and $V$ sufficiently large. For a regular space, $d_H$ always coincides with the usual
dimension $d$. In two-dimensional quantum gravity, $d_H$ need not be 2, as demonstrated by the examples of the 8q2 model coupled to Lorentzian quantum gravity with $d_H=3$ \cite{8q2LQG} and Liouville quantum gravity with $d_H=4$ \cite{ambwata}.

\begin{figure}[t]
 \begin{center}
 \subfigure{\includegraphics[scale=0.43]{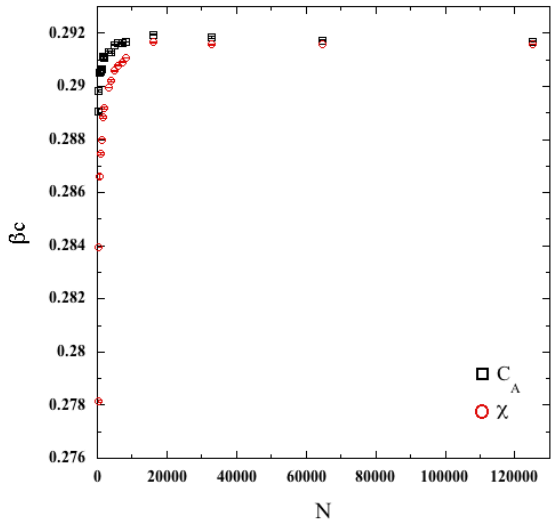}}
 \end{center}
 \caption{\small Extracting the value of the critical coupling $\beta_c$ from the finite-size scaling 
 $\beta_c(N)$ of the specific heat $C_A$ and susceptibility $\chi$, as function of the number $N$ 
 of triangles, relation (\ref{beta_c}).}
\label{fig:beta_c}
\end{figure}

\begin{figure}[t]
 \subfigure{\includegraphics[scale=0.41]{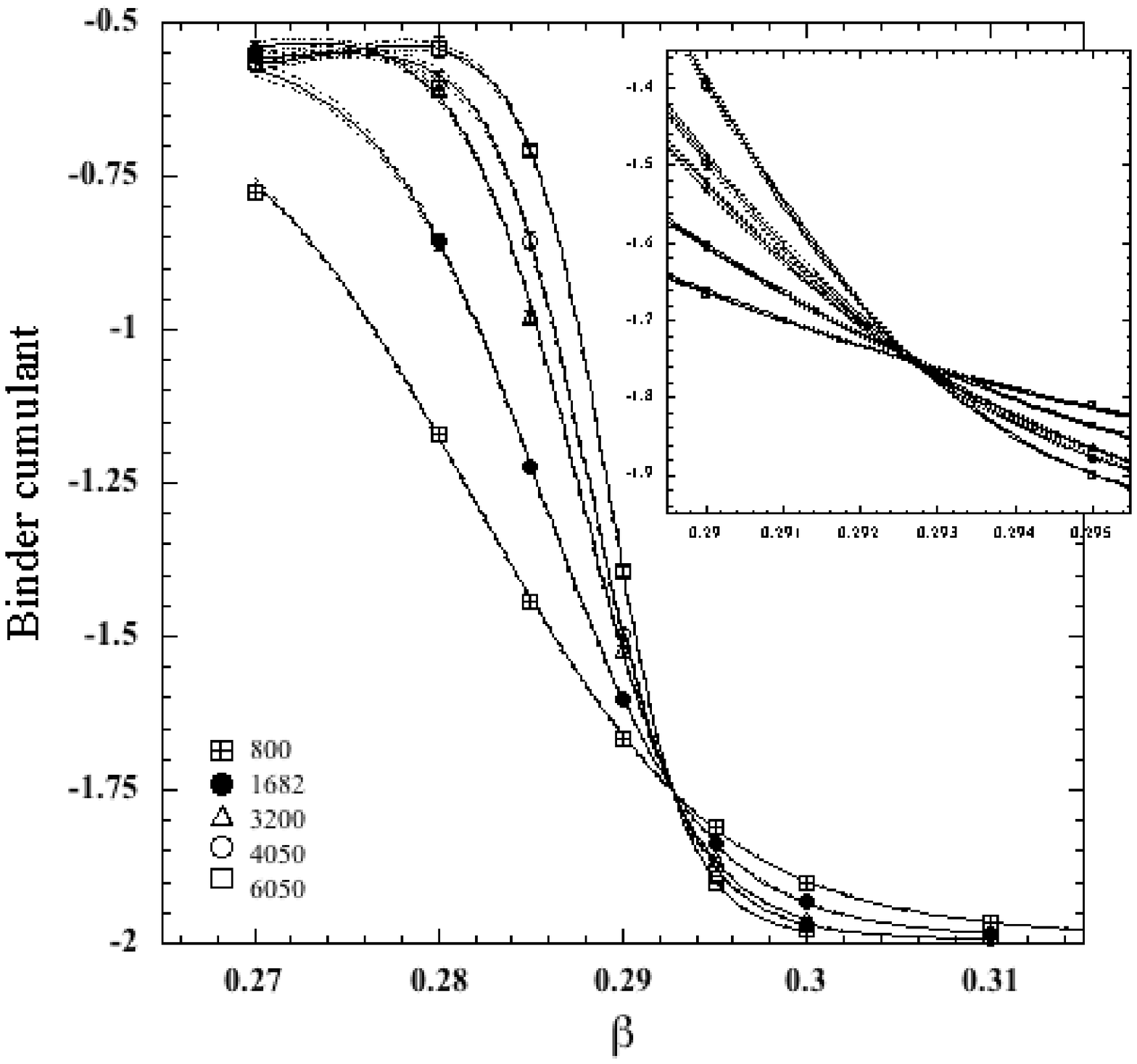}}
 \subfigure{\includegraphics[scale=0.41]{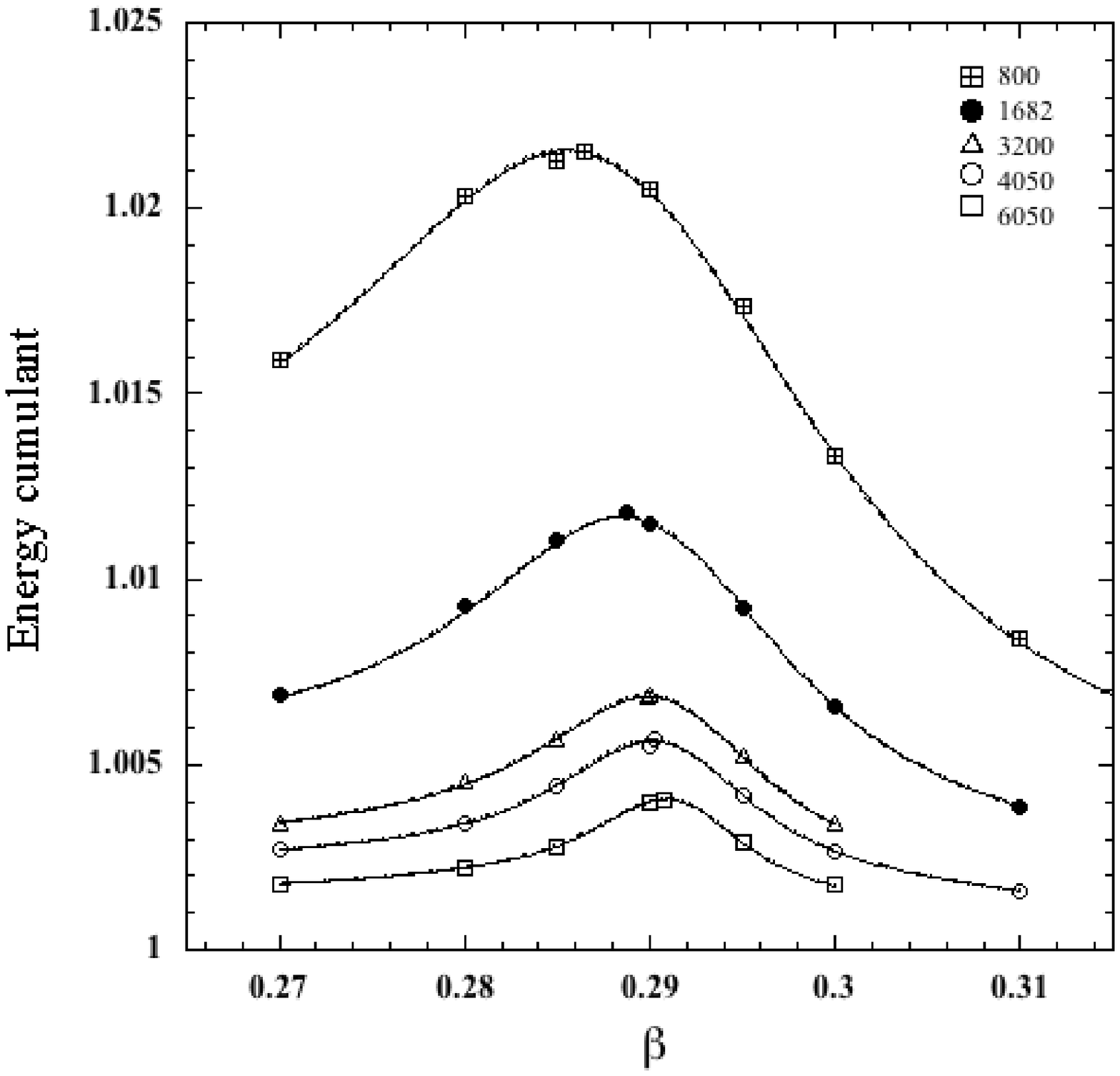}}
\caption{\small The Binder and energy cumulants (\ref{Binder}) versus $\beta$ at various 
system sizes $N$. The crossing point of the Binder cumulants (left) gives an estimate for the critical inverse temperature. The flattening of the peak in the
reduced cumulant $V_l$ (right) with increasing volume signals the presence of a second-order phase transition.} 
\label{fig:BC}
\end{figure}

As usual, since the system sizes accessible to the computer are finite, we can never observe a genuine
phase transition, but only a pseudo-critical point $\beta_c(N)$, where the linear extension of the system is of 
the order of the correlation length. This point can be determined for each volume $N$ from the resolved peaks 
of the specific heat and the magnetic susceptibility (Fig.\ \ref{peaks}). Using the extrapolating formula
\begin{equation}\label{beta_c}
\beta_c(N) \approx \beta_c + \frac{c}{N^{1/\nu d_H}},
\end{equation}
and combining the two measurements from the specific heat and magnetic susceptibility, 
we find that $\beta_c(N)$ approaches $\beta_c=0.2919(5)$ with increasing volume, see
Fig.\ \ref{fig:beta_c}. 

A useful cross-check for the location of the phase transition and its order comes from evaluating 
the Binder cumulant $BC$ and the reduced cumulant $V_l$, defined as
\begin{equation}\label{Binder}
BC = \frac{\langle m^4\rangle }{\langle m^2\rangle ^2} - 3, \qquad\qquad V_l = \frac{\langle e^4\rangle }{\langle e^2\rangle ^2}.
\end{equation}
The Binder cumulant exhibits a transition between disordered ($\beta<\beta_c$) and ordered
($\beta>\beta_c$) phase, which sharpens as the volume increases.
The intersection point of the curves $BC(\beta)$, illustrated in
Fig.\ \ref{fig:BC}, allowed us to extract an 
estimate for the location of the critical point as $\beta_c=0.2929(5)$, which is in good agreement with
the one obtained from the finite-size scaling (\ref{beta_c}).
The analogous fourth-order cumulant for the energy is the reduced cumulant
$V_l$, which is an indicator of the order of the phase transition \cite{Phase}. 
At a second-order transition, $V_l\to 1$ for all temperatures as
the volume tends to infinity, which in our case is well satisfied, as can be seen from Fig.\ \ref{fig:BC}. 
\begin{figure}[htbp]
 \begin{center}
 \subfigure{\label{fig:logSVtau1}\includegraphics[scale=0.39]{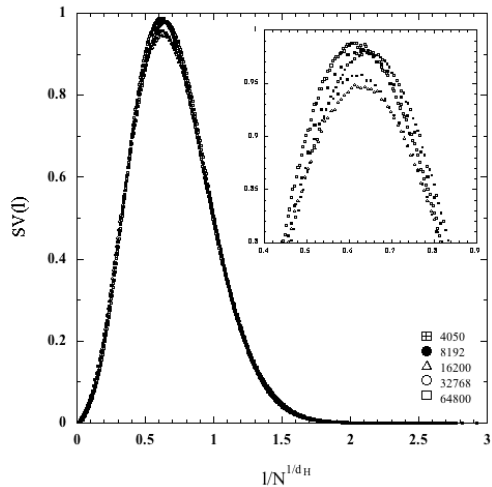}}
 \end{center}
 \caption{\small The distributions $SV_N(l)$ of the lengths $l$ of the spatial slices, suitably 
 rescaled and superposed for different volumes $N$, exhibit finite-size scaling for the optimal choice $d_H=2$. Like elsewhere, measurements at volume $N$ are taken at the pseudo-critical point $\beta_c(N)$.}
 \label{fig:SV}
\end{figure}

\section{Observables of the coupled system: results}

Having located the transition point of the Potts model, our next step is to extract the universal properties of the system at criticality in the limit as $N\to \infty$, which
characterize both the quantum geometry and the matter interacting with it. We start by looking at purely geometric observables, which take the form of certain scaling dimensions. The first of them is the large-scale Hausdorff dimension $d_H$ already introduced above. We do not a priori exclude the possibility of a different (and therefore anomalous) scaling of spatial and time-like distances, in line with the discussion in \cite{8q2LQG}. The scaling of spatial distances can be determined by measuring the 
distribution $SV_N(l)$ of spatial volumes (the lengths $l$ of the circles of constant time) in the simulation. For sufficiently large $l$ and spacetime volume $N$, we expect a universal scaling behaviour of the type
\be
SV_N(l) \sim F_S (\frac{l}{N^{1/d_H}}).
\label{volume}
\ee
With the optimal value of $d_H$, all curves $SV_N(l)$ should fall on top of each other. The optimum was
determined from a $\chi^2$-test, and the corresponding Hausdorff dimension found to be 2.000(2). 
As can be seen from Fig.\ \ref{fig:SV}, for this value finite-size scaling is satisfied for a range of ratios around the most
probable value. We have tested for a possible dependence of this result on the degree of ``elongation" of the 
spacetime histories -- as has been observed in the 8q2 system \cite{8q2LQG} -- by repeating the measurement for different ratios
$\tau:=t^2/N$, $\tau=2,3,4$, but have not found any significant dependence. 

\begin{figure}[h]
 \centering
 \includegraphics[width=.50\linewidth]{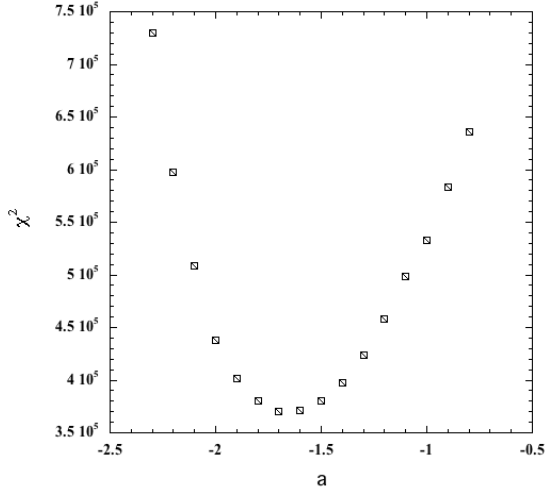}
 \caption{\small Determining the optimal additive shift $a$ in the radial variable $r$ used in the
measurement of the shell volumes, relation (\ref{nofx1q3}), by minimizing $\chi^2$.}
 \label{chi2}
\end{figure}

Another way of characterizing universal properties of the geometry, this time on short scales, is 
through the behaviour of the one-dimensional shell volumes $n_N(r)$ of spherical shells at a geodesic $r$ from a chosen reference point, as defined in \cite{shell1,shell2}. As usual, the distance $r$ is identified with link distance, and the shell volume is measured by counting the number of vertices at (integer) link distance $r$ from a given reference vertex. The shell volume is expected to exhibit a power-law behaviour of the form 
\be\label{nofx1q3}
n_N(r)\propto x^{d_h-1},\qquad x=\frac{r+a}{N^{1/d_H}}, \qquad x \ll N^{1/d_H},
\ee
defining the short-distance fractal dimension $d_h$. The shift $a$ has been introduced in order to take 
into account short-distance lattice artifacts (see \cite{shell2} for a more detailed discussion). 
For each fixed value of $a$, the joint $\chi^2$ for the extrapolated curves $n_N(x)$, with $N$ ranging over the entire volume range, is computed, leading to one of the data points in Fig.\ \ref{chi2}. At the
optimal value $a=-1.652$, where $\chi^2$ is minimal and the overlap among curves maximal, we
have extracted the value of the fractal dimension, which is given by $d_h = 1.98(1)$. 
We conclude that our measurements are compatible with $d_H=d_h=2$ and thus
all distances scale canonically in the coupled CDT-Potts system.

\begin{figure}[htp]
 \begin{center}
\subfigure{\includegraphics[scale=0.41]{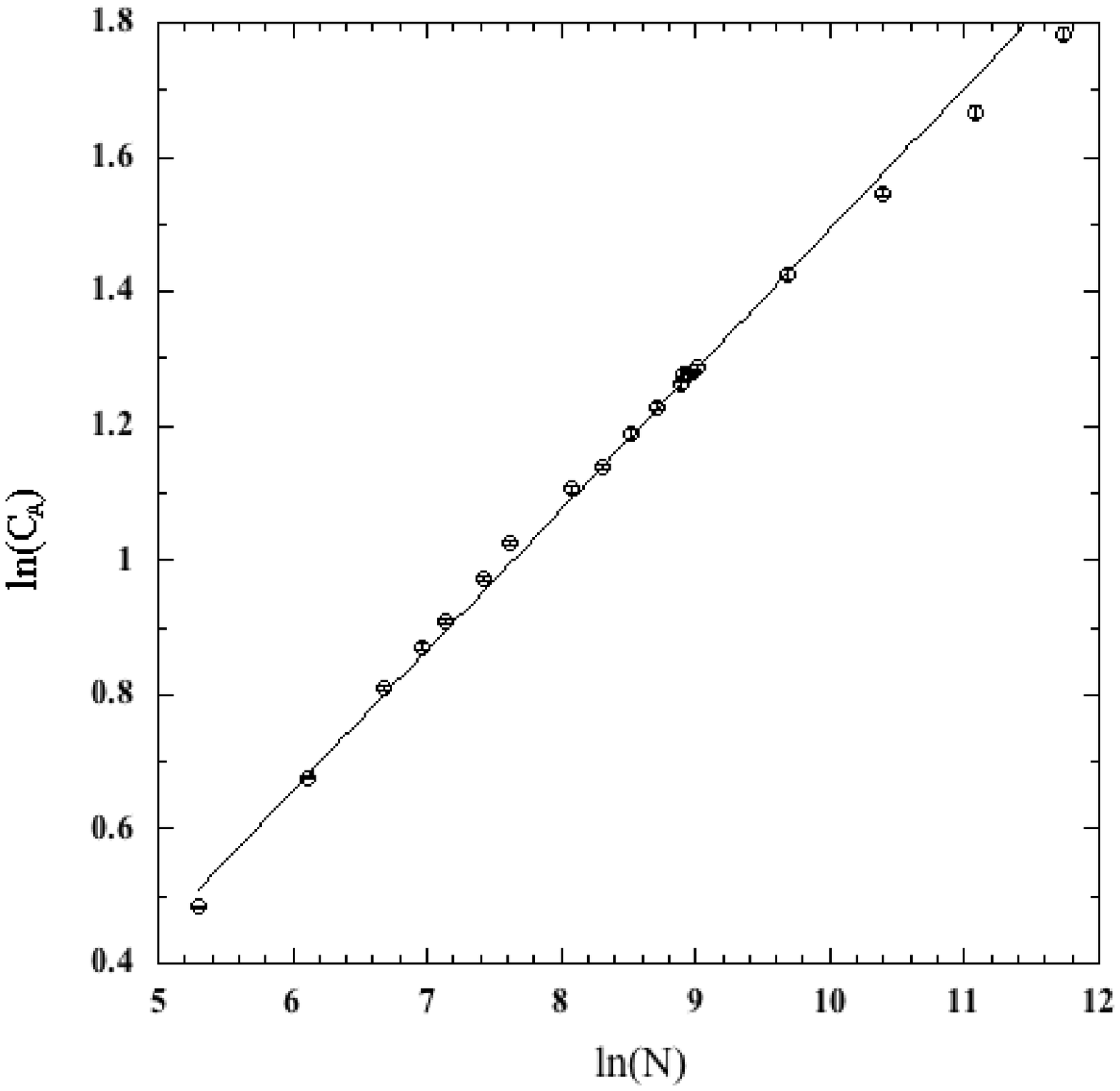}}
  \subfigure{\label{fig:susc1q3}\includegraphics[scale=0.41]{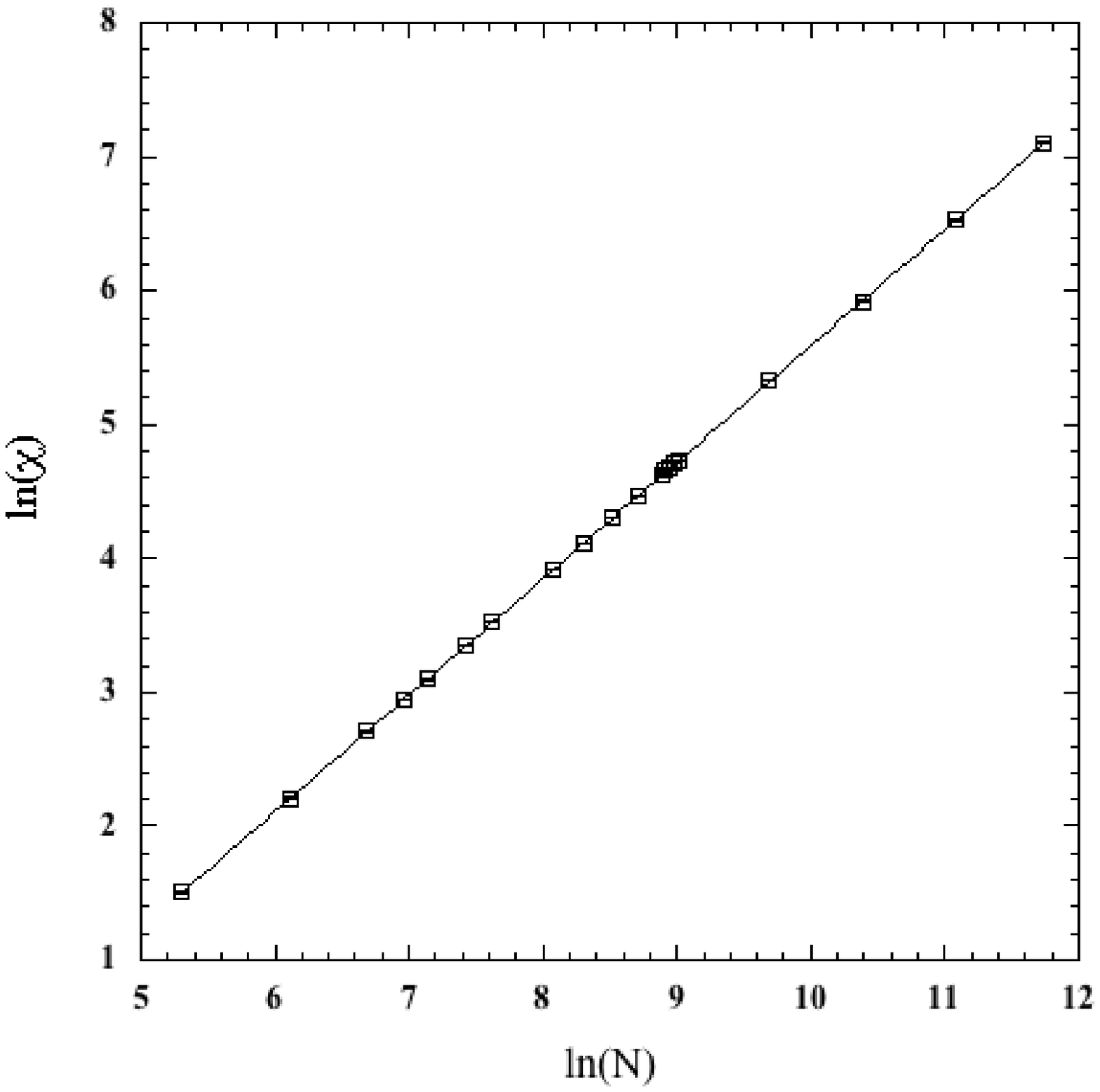}}
 \end{center}
 \caption{\small The specific heat and magnetic susceptibility, defined in (\ref{alpha}),
 (\ref{gamma}), as functions of the number $N$ of triangles, and measured at the peak of the magnetic
 susceptibility.}
 \label{fig:CAandXi}
\end{figure}
Next we turn to the matter properties at criticality, amounting to a fitting of measured data to the scaling relations 
(\ref{alpha})-(\ref{gamma}), to directly determine the quotients $\alpha/\nu d_H$, $\beta/\nu d_H$ and $\gamma/\nu d_H$. Fig.\ \ref{fig:CAandXi} illustrates some of the data taken, for the specific heat and the magnetic susceptibility. We have plotted the values at the peak of the magnetic susceptibility, as function of the volume $N$. A power scaling in $N$ is consistent with the data in all cases, and leads to the values $\alpha/\nu d_H=0.209(7)$, $\beta/\nu d_H=0.0647(3)$ and $\gamma/\nu d_H=0.8692(4)$.

\begin{figure}[hbtp]
 \centering
 \includegraphics[width=.50\linewidth]{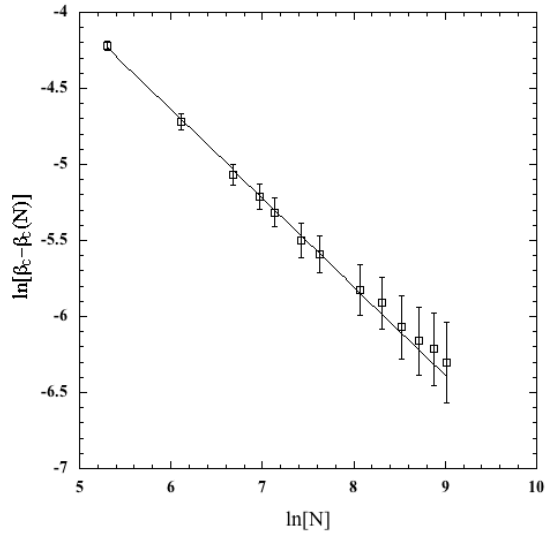}
 \caption{\small Measuring $\nu d_H$ through determining the power of $N$ in relation 
 (\ref{beta_c}) for the magnetic susceptibility.}
 \label{NDH}
\end{figure}
We have determined the combination $\nu d_H$ from relation (\ref{beta_c}) for the magnetic
susceptibility. In order to reduce
the number of fitting parameters by one, we have used the value of $\beta_c$ extracted from the
behaviour of Binder cumulants, which appears to be of very good quality. 
Using the data from the susceptibility measurement and fitting to a straight line in logarithmic scale,
we have extracted the value $\nu d_H=1.71(6)$, see Fig.\ \ref{NDH}. We have omitted the data points 
for the largest volumes, since the statistics gathered was insufficient.\footnote{We have also 
attempted an analogous fit for the specific heat, but the quality of data is poor
by comparison.}

We have collected our numerical results for the various critical exponents in Table 2, 
at the same time comparing them with those of references \cite{1q2LQG,8q2LQG} 
for the Ising model, and with the known, exact values for the same models on fixed, flat lattices. The results for the dimensions $d_H$ and $d_h$ are clear-cut: they scale canonically, which immediately excludes any resemblance with the Potts model coupled to {\it Euclidean} quantum gravity (c.f. Table \ref{table1}), as expected. Likewise, our measurement of the three matter exponents (divided by $\nu d_H$), as well as of $\nu d_H$ leaves little doubt that they are identical with the corresponding
values on regular, fixed lattices. A quick cross-check reveals that the scaling identities (\ref{hyper}) are satisfied at a level of about 1\%.   

\begin{table}[htb]
 \centering
 \begin{tabular}{|c|c|c|c|c|c|}\toprule
 \multicolumn{6}{|c|}{\bf Potts models on flat and dynamical LQG lattices}   \\ \cmidrule(r){2-6}
          & {\bf 1q2, flat} & {\bf 1q2, LQG} & {\bf 1q3, flat} & {\bf 1q3, LQG} & {\bf 8q2, LQG}          \\
  {\bf Indices} &                  & reference \cite{1q2LQG} &    & this paper    & reference \cite{8q2LQG}   \\ \midrule
      $c$       &        1/2         &      1/2         &    4/5    &  4/5 &    4             \\ \midrule
$\alpha/\nu d_H$&    0              &    0.0861(1) & 0.2    &  0.209(7) &                    \\
$\beta/\nu d_H$ &     0.0625    &     0.070(1) &  0.0667    & 0.0647(3) &                    \\
$\gamma/\nu d_H$&   0.875    &     0.883(1) & 0.8667    & 0.8692(4) &     0.85(1)        \\ 
$\nu d_H$    &     2             &     1.97(6)      & 1.6667 & 1.71(6)  &     1.85(1)        \\\midrule
     $d_H$      &        2         &     2.00(4)   & 2   & 2.000(2)  &     3.07(9)        \\
     $d_h$      &        2         &        2.00(5)  &  2         &  1.98(1)  &     2.1(2)       \\ \bottomrule
\end{tabular}
\caption{\small Comparing critical exponents of the two- and three-state Potts models, on fixed, flat 
 lattices and coupled to Lorentzian quantum gravity in the form of an ensemble of causal dynamical triangulations. Like in the case of the Ising model (1q2) studied previously, our present investigation provides strong evidence that also for the three-state Potts model the critical exponents on flat and on dynamical CDT lattices coincide.}
\label{table2}
\end{table}

\section{Conclusion}

We have investigated two-dimensional Lorentzian quantum gravity, in the form of an ensemble of causal dynamical triangulations, coupled to a single copy of the three-state Potts model. This spin model is of particular interest, because its critical coefficients are known exactly both on fixed, regular lattices and when coupled to Euclidean dynamical triangulations, and thus can be compared to, and because it has a positive specific-heat coefficient $\alpha$, which may make its universal properties susceptible to random disorder. We were therefore particularly interested in the effect of the geometric disorder inherent in the gravity model on the critical behaviour of the spin system. Using Monte Carlo methods and finite-size scaling techniques, we found that the universal properties of the gravitational sector, expressed in terms of dynamical critical dimensions were unchanged from the pure gravitational model. 
With regard to the spin sector, we observed the customary second-order transition, 
and critical exponents for the specific heat, magnetization and magnetic susceptibility were found to be in good agreement with the corresponding values on fixed, regular lattices. 
For $\nu d_H$ ($\nu$ being the critical exponent of the spin-spin correlation length and $d_H$ the large-scale Hausdorff dimension) the data quality is slightly inferior (mirroring a similar behaviour as
found for the Ising model), but agreement is still good. Overall, the coincidence with the fixed-lattice
critical exponents for the three-state Potts model is comparable to and even slightly better than what was found previously for the
Ising model \cite{1q2LQG}, as can be seen from Table 2. In summary, in contrast to
what may have been expected from the effects of disorder for a system with positive specific
heat, we have found compelling numerical evidence that the strong geometric disorder
implied by the Lorentzian gravity ensemble does not alter the universality class of a three-state
Potts model coupled to it. -- The Potts spins are shaken, but not stirred!

\vspace{0.8cm}

\noindent \textbf{\it Acknowledgements.} RL thanks Des Johnston for discussion. All authors acknowledge the support of 
ENRAGE (European Network on
Random Geometry), a Marie Curie Research Training Network in the
European Community's Sixth Framework Programme, network contract
MRTN-CT-2004-005616. RL acknowledges
support by the Netherlands
Organisation for Scientific Research (NWO) under their VICI
program.

\vspace{1.2cm}


\begin{thebibliography}{99}

\bibitem{4deu}
P.~Bialas, Z.~Burda, A.~Krzywicki and B.~Petersson:
{\it Focusing on the fixed point of 4d simplicial gravity,}
Nucl.\ Phys.\ B\ 472 (1996) 293-308 [hep-lat/9601024];
P.~Bialas, Z.~Burda, B.~Petersson and J.~Tabaczek:
{\it Appearance of mother universe and singular vertices in random
geometries,}
Nucl.\ Phys.\ B\ 495 (1997) 463-476 [hep-lat/9608030].

\bibitem{ajl-rec} 
J.\ Ambj\o rn, J.\ Jurkiewicz and R.\ Loll: {\it Reconstructing the universe},
Phys.\ Rev.\ D\ 72 (2005) 064014  [hep-th/0505154].

\bibitem{ajl-prl}
J.~Ambj\o rn, J.~Jurkiewicz and R.~Loll:
{\it Emergence of a 4D world from causal quantum gravity},
Phys.\ Rev.\ Lett.\ 93 (2004) 131301 [hep-th/0404156];
{\it Semiclassical universe from first principles},
Phys.\ Lett.\ B 607 (2005) 205-213
[hep-th/0411152].

\bibitem{agjl} 
J.~Ambj\o rn, A.\ G\"orlich, J.~Jurkiewicz and R.~Loll:
{\it Planckian birth of the quantum de Sitter universe},
Phys.\ Rev.\ Lett.\ 100 (2008) 091304 [0712.2485, hep-th].

\bibitem{al} J.\ Ambj\o rn and R.\ Loll:
{\it Non-perturbative Lorentzian quantum gravity, causality and
topology change},
Nucl.\ Phys.\ B\ 536 (1998) 407-434 [hep-th/9805108].

\bibitem{2d}
J.~Ambj\o rn, R.~Loll, J.~L.~Nielsen and J.~Rolf:
{\it Euclidean and Lorentzian quantum gravity: Lessons from two dimensions},
Chaos Solitons Fractals\ 10 (1999) 177-195
[hep-th/9806241];
J.~Ambj\o rn, J.~Jurkiewicz and R.~Loll:
{\it Lorentzian and Euclidean quantum gravity: Analytical and numerical
results},
in Proceedings of M-Theory and Quantum Geometry, 1999 NATO Advanced Study Institute, Akureyri, Island, 
eds. L. Thorlacius et al. (Kluwer, 2000) 382-449
[hep-th/0001124].

\bibitem{ajwz}
J.~Ambj\o rn, R.~Janik, W.~Westra and S.~Zohren:
{\it The emergence of background geometry from quantum fluctuations},
Phys.\ Lett.\  B\ 641 (2006) 94-98
[gr-qc/0607013].

\bibitem{alwz}
J.~Ambj\o rn, R.~Loll, W.~Westra and S.~Zohren:
{\it Putting a cap on causality violations in CDT},
JHEP 0712 (2007) 017 [0709.2784, gr-qc].
  
\bibitem{liouville}
 P.~Di Francesco, P.H.~Ginsparg and J.~Zinn-Justin:
 {\it 2-D Gravity and random matrices},
 Phys.\ Rept.\  254 (1995) 1-133 [hep-th/9306153].

\bibitem{ad}
J.~Ambj\o rn and B.~Durhuus:
{\it Regularized bosonic strings need extrinsic curvature},
Phys.\ Lett.\  B\ 188 (1987) 253-257.

\bibitem{david}
F.~David:
{\it A scenario for the c $ >$ 1 barrier in non-critical bosonic strings},
Nucl.\ Phys.\  B\ 487 (1997) 633-649 [hep-th/9610037].
 
\bibitem{1q2LQG} 
J.~Ambj\o rn, K.N.~Anagnostopoulos and R.~Loll:
{\it A new perspective on matter coupling in 2d quantum gravity},
Phys.\ Rev.\  D\ 60 (1999) 104035 [hep-th/9904012].

\bibitem{8q2LQG} 
J.~Ambj\o rn, K.N.~Anagnostopoulos and R.~Loll:
{\it Crossing the c $= $1 barrier in 2d Lorentzian quantum gravity},
Phys.\ Rev.\  D\ 61 (2000) 044010 [hep-lat/9909129].

\bibitem{jajo} 
W.~Janke and D.A.~Johnston:
{\it Ising and Potts models on quenched random gravity graphs},
Nucl.\ Phys.\  B\ 578 (2000) 681-698 [hep-lat/9907026].
 
\bibitem{baillie}
C.F.~Baillie and D.A.~Johnston:
{\it A Numerical test of KPZ scaling: Potts models coupled to two-dimensional
quantum gravity},
Mod.\ Phys.\ Lett.\  A\ 7 (1992) 1519-1534 [hep-lat/9204002].

\bibitem{johnston}
D.A.~Johnston:
{\it Zero Potts models coupled to two-dimensional quantum gravity},
Phys.\ Lett.\  B\ 277 (1992) 405-410.

\bibitem{harris} A.B.~Harris:
{\it Effect of random defects on the critical behaviour of Ising models},
J.~Phys.\ C7 (1974) 1671-1692.

\bibitem{luck} J.M.\ Luck:
{\it A classification of critical phenomena on quasi-crystals and other 
aperiodic structures},
Europhys.\ Lett.\ 24 (1993) 359-364.

\bibitem{jawei} W.\ Janke and M.\ Weigel: 
{\it The Harris-Luck criterion for random lattices},
Phys.\ Rev.\ B\ 69 (2004) 144208 [cond-mat/0310269].

\bibitem{ising} D.\ Benedetti and R.\ Loll:
{\it Unexpected spin-off from quantum gravity},
Physica A 377 (2007) 373-380 [hep-lat/0603013];
{\it Quantum gravity and matter: counting graphs on causal dynamical
triangulations},
Gen.\ Rel.\ Grav.\ 39 (2007) 863-898 [gr-qc/0611075].

\bibitem{christ} N.H.\ Christ, R.\ Friedberg and T.D.\ Lee: 
{\it Random lattice field theory}, Nucl.\ Phys.\ B\ 202 (1982) 89-125; 
{\it Gauge theory on a random lattice},
Nucl.\ Phys.\ B\ 210 (1982) 310-336;
{\it Weights of links and plaquettes in a random lattice}
Nucl.\ Phys.\ B\ 210 (1982) 337-346.

\bibitem{matrix}
J.~Ambj\o rn, R.~Loll, Y.~Watabiki, W.~Westra and S.~Zohren:
{\it A matrix model for 2D quantum gravity defined by causal dynamical
triangulations} [0804.0252, hep-th].

\bibitem{3d4d} J.\ Ambj\o rn, J.\ Jurkiewicz and R.\ Loll:
{\it Dynamically triangulating Lorentzian quantum gravity},
Nucl.\ Phys.\ B\ 610 (2001) 347-382 [hep-th/0105267].

\bibitem{DT-book} J.\ Ambj\o rn, B.\ Durhuus and T.\ Jonsson: 
{\it Quantum geometry},
Cambridge University Press (1997).

\bibitem{ambwata} J.~Ambj\o rn and Y.~Watabiki:
{\it Scaling in quantum gravity},
Nucl.\ Phys.\  B\ 445 (1995) 129-144 [hep-th/9501049].

\bibitem{Phase} A.M.~Ferrenberg and R.H.~Swendsen: 
{\it New Monte Carlo technique for studying phase transitions},
Phys.~Rev.~Lett.~61 (1988) 2635-2638.

\bibitem{shell1} 
S.~Catterall, G.~Thorleifsson, M.J.~Bowick and V.~John:
{\it Scaling and the fractal geometry of two-dimensional quantum gravity},
Phys.\ Lett.\  B\ 354 (1995) 58-68 [hep-lat/9504009].

\bibitem{shell2}
J.~Ambj\o rn, J.~Jurkiewicz and Y.~Watabiki:
{\it On the fractal structure of two-dimensional quantum gravity},
Nucl.\ Phys.\  B\ 454 (1995) 313-342 [hep-lat/9507014];
J.~Ambj\o rn, K.N.~Anagnostopoulos, U.~Magnea and G.~Thorleifsson:
{\it Geometrical interpretation of the KPZ exponents},
Phys.\ Lett.\  B\ 388 (1996) 713-719 [hep-lat/9606012].


\end{thebibliography}
\end{document}